\documentclass[aps,prl,reprint,superscriptaddress]{revtex4-1}
\usepackage{graphicx}

\usepackage{color}
\usepackage{enumerate}
\usepackage{amsmath}\usepackage{amssymb}\usepackage{stmaryrd}
\usepackage{multirow}
\usepackage{float}
\usepackage{longtable,booktabs}
\usepackage{color}\usepackage{enumerate}
\usepackage{amsmath}\usepackage{amssymb}\usepackage{stmaryrd}
\usepackage{multirow}
\usepackage{float}
\usepackage[toc,page]{appendix}
\usepackage{longtable,booktabs}
 \usepackage{epsfig}
\usepackage{graphicx} 
\usepackage{soul}
\usepackage{bm}

\usepackage{psfrag} 
\usepackage{slashed}
\usepackage{amsmath,amssymb} 
\usepackage{colordvi} 
\usepackage{hyperref}
\usepackage{setspace}
\usepackage[all]{hypcap}
\usepackage{natbib}
\usepackage{subfigure}

\hypersetup{
    bookmarks=true,         % show bookmarks bar?
    unicode=false,          % non-Latin characters in AcrobatÕs bookmarks
    pdftoolbar=true,        % show AcrobatÕs toolbar?
    pdfmenubar=true,        % show AcrobatÕs menu?
    pdffitwindow=false,     % window fit to page when opened
    pdfstartview={FitH},    % fits the width of the page to the window
    pdftitle={My title},    % title
    pdfauthor={Author},     % author
    pdfsubject={Subject},   % subject of the document
    pdfcreator={Creator},   % creator of the document
    pdfproducer={Producer}, % producer of the document
    pdfkeywords={keyword1} {key2} {key3}, % list of keywords
    pdfnewwindow=true,      % links in new window
    colorlinks=true,       % false: boxed links; true: colored links
    linkcolor=red,          % color of internal links
    citecolor=green,        % color of links to bibliography
    filecolor=magenta,      % color of file links
    urlcolor=black           % color of external links
}
\def\be{\begin{equation}}
\def\bea{\begin{eqnarray}}
\def\eea{\end{eqnarray}}
\def\ee{\end{equation}}
\def\del{\partial}

\def\r{{\bf r}}

\DeclareMathOperator{\Imm}{Im}
\renewcommand{\[}{\left [}
\renewcommand{\]}{\right ]}
\renewcommand{\(}{\left (}
\renewcommand{\)}{\right )}

\begin{document}

\title{Topological surface superconductivity in doped Weyl loop materials}
\author{Yuxuan Wang}
\affiliation{Department of Physics and Institute for Condensed Matter Theory, University of Illinois at Urbana-Champaign, Urbana, Illinois 61801, USA}
\author{Rahul M. Nandkishore}
\affiliation{Department of Physics and Center for Theory of Quantum Matter,
Univeristy of Colorado, Boulder, Colorado 80309, USA}

\begin{abstract}
We study surface superconductivity involving the `drumhead' surface states of (doped) Weyl loop materials. The leading weak coupling instability in the bulk is toward a chiral superconducting order, which fully gaps the Fermi surface. In this state the surface also becomes superconducting, with $p+ip$ symmetry. We show that the surface SC state is ``topological" as long as it is fully gapped, and the system traps Majoarana modes wherever a vortex line enters or exits the bulk. In contrast to true 2D $p+ip$ superconductors, these Majorana zero modes arise even in the ``strong pairing" regime where the chemical potential is entirely above/below the drumhead. We also consider conventional $s$-wave pairing, and show that in this case the surface hosts a flat band of charge neutral Majorana fermions, whose momentum range is given by the projection of the bulk Fermi surface.
Weyl loop materials thus provide access to new forms of topological superconductivity.
\end{abstract}

%\pacs{}
\maketitle
%\tableofcontents

Weyl loop systems are a new class of topological semimetals that has recently been proposed \cite{Carteretal, ChenLuKee, Schafferetal, KimWiederKaneRappe, Mullenetal, aletFu, aletBernevig, aletHu} and observed \cite{Cava, NeupertHasan}. These are three-dimensional materials which at a particular filling display a one dimensional ring Fermi surface in the bulk (which becomes toroidal upon doping), with massless Dirac-like quasiparticles. These materials constitute a new class of electronic system that is intermediate between Weyl semimetals (where the Fermi surface is pointlike), and ordinary three dimensional metals (with a two dimensional Fermi surface). As such, they can support a qualitatively new phenomenology which is exciting considerable interest \cite{RhimKim,Mullenetal, LiuBalents}. 

While most research on Weyl loop materials has focused on the {\it non-interacting} materials, these systems also provide a new playground for investigating the {\it correlated states} that may arise through weak coupling instabilities. For example, Ref.\ \cite{WeylLoopSC} used general symmetry arguments to establish that Weyl loop materials are natural candidates to host exotic forms of superconductivity (SC) in the bulk, with the leading weak coupling instability being to a fully gapped chiral superconducting state in three dimensions, which has never before been observed. This conclusion was also borne out by detailed microscopic calculations within a renormalization group formalism in \cite{SurNandkishore}. Separately, it has been pointed out that Weyl loop materials host `drumhead' surface states \cite{BurkovHookBalents} {(see also \cite{YZ1, YZ2, YZ3})} with a large density of states, which could naturally support high temperature surface SC \cite{volovik2}. However, a detailed analysis of the symmetry and topology of surface superconductivity in these materials has never yet been performed~\cite{footnote1}. 

In this Letter we provide the first systematic analysis of the symmetry and topology structure of surface superconductivity in Weyl loop materials, as well as its interplay with bulk superconductivity and the superconducting proximity effect. We begin by reviewing the structure of bulk SC before turning to surface SC. We discuss ``conventional" $s$-wave state, and show that the surface state is immune to proximity effect from the $s$-wave order. However bulk $s$-wave pairing induces gapless surface SC, with a new drumhead of surface states made out of charge neutral Majorana fermions. We then discuss surface SC in the presence of bulk chiral $p$-wave supercondivity, which is \cite{WeylLoopSC, SurNandkishore}  the only fully gapped bulk state. In this case we show that the surface develops $p+ip$ chiral superconducting order, and is fully gapped except for some special cases where the drumhead band is at zero energy at an time-reversal invariant (TRI) momentum. In analogy with the two-dimensional (2D) $p+ip$ superconductor \cite{moore-read,Anyon2}, it is tempting to classify the surface SC with a Fermi surface (FS) of the drumhead band as topological and the opposite case as topologically trivial. However, we show that this is not the case --  as long as the surface SC is gapped, it is {\it topological} and traps Majorana zero modes (MZM) where a vortex line enters or exits the bulk. This is reminiscent of \cite{FuKane}. Here the distinction with 2D $p+ip$ SC arises because the drumhead band does not cover the full 2D Brillouin zone (BZ).
We establish this by means of analytic arguments based on adiabatic continuation, and also verify this  numerically. Weyl loop materials thus provide access to new forms of topological (surface) superconductivity. 

\begin{figure}
\includegraphics[width=0.7\columnwidth]{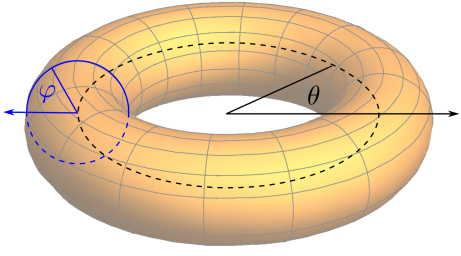}
\caption{The torus-shaped Fermi surface of a Weyl loop metal.}
\label{torus}
\end{figure}

{\it Model.---}
We consider a two-band model,
\begin{align}
\mathcal{H}({\bf k})=&\sigma^x(6-t_1-2\cos k_x-2\cos k_y-2\cos k_z)\nonumber\\
&+2t_2\sigma^y\sin(k_z)-\mu,
\label{h}
\end{align}
where we have factored out an overall energy scale, and the matrices $\sigma^x$ and $\sigma^y$ (hereafter ``spin"), can either be in $SU(2)$ spin space, or simply in a two-band subspace of a multi-band system.
The first two terms describe a line node in the dispersion, given by $3-t_1/2-\cos k_x-\cos k_y=0$ and $k_z=0$.
This model has a $\mathcal{TI}$ symmetry~\cite{BurkovHookBalents,srinidhi}, which is a product of time-reversal $\mathcal{T}=\sigma^zK$ and spatial inversion $\mathcal{I}=\sigma^y$, such that
$
\mathcal{H}({\bf k})={(\mathcal{TI})}\mathcal{H}({\bf k}){(\mathcal{TI})^{-1}}.
$ Such a composite $\mathcal{TI}$ symmetry ensures that the line node cannot be locally gapped out, since $\mathcal{TI}$ symmetry dictates the Berry curvature to be zero everywhere (except for at singular points, i.e., the line node), and gapping out the line nodes would smear out the Berry curvature in $k$-space. 
 The last term is a chemical potential that gives rise to a non-degenenrate Fermi surface surrounding the nodal line, which forms a torus.
The dispersion in the continuum limit can be written as
\begin{align}
\mathcal{H}=&\sigma^x(k_\|^2-t_1)+2t_2\sigma^y k_z -\mu,
\end{align}
where $k_\|^2=k_x^2+k_y^2$.
At a given polar angle $\theta$ (defined by $k_y/k_x=\tan\theta$), for a small $\mu$, the cross-section of the FS is given by two ellipses at $t_1(k_\|\pm \sqrt{t_1})^2+t_2k_z^2=\mu^2$. Without loss of generality we set $t_1=t_2$ and parametrize this circle by another angle $\varphi$, as shown in Fig.\ \ref{torus}, such that the linearized Hamiltonian takes the form $H = K (\cos \varphi \sigma^x + \sin \varphi \sigma^y)$, where $K\cos \varphi = k_\| - \sqrt{t_1}$ and $K \sin \varphi = t_1 k_z$. Note that inversion takes $\theta \rightarrow \theta + \pi$ and $\varphi \rightarrow - \varphi$ with this parametrization. 

{\it Superconductivity in the bulk.---}
We first consider the superconducting order parameter in the odd-parity channels that couples to fermions via
\begin{align}
H^o_\Delta&=\Delta_{\alpha\beta} \psi^\dagger_\alpha({\bf k}) \psi^\dagger_{\beta}(-{\bf k})\nonumber\\
&=\Delta \psi^\dagger({\bf k}) ({\bf d(k)}\cdot {\bm \sigma})i\sigma^y(\psi^\dagger)^T(-{\bf k}).
\end{align}
Fermionic statistics require $\bf d(k)$ to be odd in $\bf k$. From the torus configuration of the FS, we expect that energetically the leading SC instability is toward an order parameter with ${\bf d(k)}\propto e^{i\theta}\tilde {\bf d}({\bf k})$, where $\tilde {\bf d}({\bf k})$ is an even function. We focus on the cases where $\tilde {\bf d}({\bf k})$ is a constant {on the FS}, and is pointed toward $x,y$, or $z$ directions. Because of the nontrivial spin-texture, the SC order parameter projects onto the low energy fermions with nontrivial ``projective form factors"~\cite{wang-ye}. To this end, we note that the spinor structure of a low energy fermion  (conduction band) on the FS is given by $c^\dagger({\bf k})=\sum_{\alpha=\uparrow,\downarrow} \xi_\alpha \psi^\dagger_\alpha({\bf k})$, where
$\xi=[\exp (i\varphi/2),\exp(-i\varphi/2)]^T/\sqrt2$. When projected onto the low energy fermions, the pairing vertex becomes $\bar H_\Delta=\Delta \chi({\bf k}) c^\dagger({\bf k}) c^\dagger({{-\bf k}})$,
where $\chi({\bf k})\equiv \xi^\dagger({\bf k}) ({\bf d}\cdot {\bm \sigma})i\sigma^y\xi^*(-{\bf k})$.  The projective form factors (on the conduction band) for $\tilde{\bf d}=d\hat x, \hat y , \hat z$ are given by 
\begin{align}
 \chi_x(\theta,\varphi)=&-\xi_\alpha^\dagger(\varphi)\sigma^z_{\alpha\beta}\xi_\beta^*(-\varphi)e^{i\theta}=0,\nonumber\\
\chi_y(\theta,\varphi)=&~i\xi_\alpha^\dagger(\varphi)\delta_{\alpha\beta}\xi_\beta^*(-\varphi)e^{i\theta}=ie^{i\theta},\nonumber\\
 \chi_z(\theta,\varphi)=&-\xi_\alpha^\dagger(\varphi)\sigma^x_{\alpha\beta}\xi_\beta^*(-\varphi)e^{i\theta}=-\cos\varphi\,e^{i\theta},
\end{align}
%where we have used the fact that ${\bf k}$ and ${\bf -k}$ corresponds to $(\theta,\varphi)$ and $(\pi+\theta,\pi-\varphi)$. 
From this we see that in the odd-parity channel, the only order that fully gaps the FS is the one with $\tilde{\bf d}=d\hat y$, while for $\tilde{\bf d}=d\hat x, d\hat z$ the SC order parameter either has no FS component, or leaves a nodal line. This is consistent with Ref.\ \cite{WeylLoopSC}, which pointed out that $\tilde{\bf d}=d\hat y$ is the only odd parity channel to be fully gapped and to involve only intra-band pairing. 
Since inter-band pairing does not contribute to the pairing instability at weak coupling, and since condensation energy is maximized by a full gap, the $d \hat y$ state is expected to be the leading odd parity instability -- as confirmed by detailed renormalization group calculations in \cite{SurNandkishore}. Or course, odd parity superconductivity needs to be ``seeded" by a bare attraction in an odd angular momentum channel (which could come e.g.\ from the Kohn-Luttinger mechanism \cite{KL, NandkishoreThomaleChubukov}). 
  
\begin{figure}
\includegraphics[width=\columnwidth]{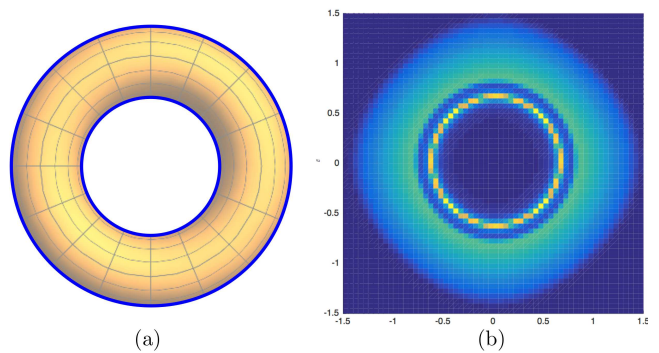}
\caption{Panel (a): The line node of the $s$-wave state shown on the torus FS (in blue), which corresponds to $\varphi=\pi/2$ and $\varphi=3\pi/2$. Panel (b): Simulation of the ARPES data of the surface states, showing the coexistence of the Majorana flat band and the drumhead FS. {The color code denotes the surface fermionic spectral weight at zero energy, given by $\Imm{G(\omega=0,{\bf k}_{xy},z=0)}$. In the Green's function $G$ we have added a very small imaginary self-energy for illustration purposes.} }
\label{mflatband}
\end{figure}

We also consider the SC order in the conventional $s$-wave channel, which couples to fermions via
\begin{align}
H^s_\Delta&=\bar\Delta_{\alpha\beta} \psi^\dagger_\alpha({\bf k}) \psi^\dagger_{\beta}(-{\bf k})=\bar\Delta \psi^\dagger({\bf k}) i\sigma^y(\psi^\dagger)^T(-{\bf k}).
\end{align}
Its projective form factor can be obtained by 
  \begin{align}
 \chi_s(\varphi)=&i\xi_\alpha^\dagger(\varphi)\sigma^y_{\alpha\beta}\xi_\beta^*(-\varphi)=-i\sin\varphi.
 \end{align}
 Thus in the $s$-wave superconducting states there are {\it two} nodal lines at $\varphi=0$ and $\varphi=\pi$ [see Fig.\ \ref{mflatband}(a)].  The existence of the two nodal lines tends to strongly suppress $T_c$ for even parity order. However attractive interactions in the even-parity channel can be generated by more conventional mechanisms (e.g.\ phonons), and thus we will discuss $s$-wave order too. {Note that $s$ and $p$ wave orders do not couple because they transform differently under $\pi$ rotations in the $xy$ plane, and also have opposite mirror eigenvalues (where mirror symmetry is generated in the projected model by $\varphi \rightarrow - \varphi$ and in the full model by $k_z \rightarrow -k_z$ and conjugation by $\sigma_1$).} A full comparative analysis of the respective $T_c$'s in the different channels requires a detailed knowledge of microscopic interactions which is beyond the scope of the current work.

%For ${\bf d}=d\hat x$, the bulk is not gapped; for ${\bf d}=d\hat y$, the bulk has two line nodes at $\varphi=0,\pi$; for ${\bf d}=d\hat z$, the bulk is fully gapped.
%Meanwhile, there exist a drumhead band on each surface, whose spin is fully polarized along $\sigma^y$. for ${\bf d}=d\hat x$, the surface FS is gapped; for ${\bf d}=d\hat y$, the surface FS remains gapless; for ${\bf d}=d\hat z$ the surface is fully gapped, as well as the bulk. These are confirmed by numerics.
%
%Using the fact that for ${\bf k}=(\theta,\varphi)$, $-{\bf k}=(\pi+\theta,\pi-\varphi)$
%

{\it Surface SC.---} We now turn to surface SC. 
For a Weyl loop band structure, there exist  ``drumhead" bands localized on the 2D surface. Unlike true 2D energy bands, these do not extend over the full surface BZ, but are confined in a momentum range given by the projection of the bulk line node. 
The drumhead bands can be obtained by treating the Hamiltonian (\ref{h}) at any given $(k_x,k_y)$ point as a one-dimensional subsystem along $z$ direction, and computing its end states. Recall that Wilson loop around a point on the line node has a Berry phase of $\pi$ (since the spin winds by $2\pi$). One can deform the Wilson loop into two paths in $z$ direction across the BZ, which correspond to the polarizations for two 1D subsystems inside and outside the line node. The $\pi$ Berry phase around the original Wilson loop indicates that the polarization of the two subsystems differs by $eL/2$~\cite{SSH,srinidhi} ($L$ is the system size),  indicating degenerate end states for subsystems either inside or outside the line node.
For our model, we compute the surface states using  open boundary conditions in the $z$ direction~\cite{SSH,Jackiw,Martin2008}. We can effectively treat the physical surface as an interface with a trivial insulator with $H=M_0 \sigma^x$ where $M_0>0$. The surface states are given by $\sigma^z\Psi=\pm \Psi$ for the top and bottom surface respectively. The two surface states are degenerate, and their dispersion is  given by $E=-\mu$.

For our purposes it is interesting to allow for a dispersion in the last term of Eq.\ \eqref{h}, i.e., $\mu=\mu(k_x,k_y)$, which preserves the $\mathcal{TI}$ symmetry (and also has the same rotation symmetry in the $xy$ plane as the bulk). In this case the drumhead band can have a Fermi surface of its own. For a small dispersion in $\mu$, the density of states on the surface is large, and the surface bands are also unstable towards SC in the presence of interactions \cite{volovik2}. However, since the low energy fermions are fully spin-polarized, $s$-wave order cannot develop. Thus the drumhead band is  immune to an external proximity effect from an $s$-wave SC. {In contrast in the bulk projection onto one band generates a spin texture, so bulk s-wave pairing is allowed.} In the event of bulk $s$-wave pairing, the drumhead band remains gapless. However, the existence of two new nodal lines in the bulk band structure coming from the SC order induces a new {\it flat} band on each surface, which cannot disperse, because of particle-hole symmetry in Nambu space. This flat band can again be viewed as a result of $\pi$ Berry flux around each nodal line~\cite{SM},
similar to the drumhead band. This band is formed by charge-neutral Majorana fermions, and can be viewed as the stacking of Majorana end states from 1D topological superconductors. We show in Fig.\ \ref{mflatband}(b) the simulation of surface ARPES data [momentum distribution curves at energy $\epsilon=0$], where both the Majorana flat band and the gapless  drumhead FS are clearly visible. 

We now discuss odd parity bulk superconductivity in the $\tilde{\bf d}=d\hat y$ channel that fully gaps out the bulk. The SC form factor on the top surface is given by 
\begin{align}
\chi_y^s(\theta,\varphi)=i(\xi^s)^\dagger(\xi^s)^*e^{i\theta}=ie^{i\theta},
\end{align}
where $\xi^s\equiv (1,0)^T$ is the spinor structure for the top surface state. From this, we find that the {\it same} superconducting order that fully gaps out the bulk also fully gaps the surface states with $p+ip$ symmetry.
Even for cases where the surface band does not cross chemical potential, the surface becomes superconducting due to proximity effect from the bulk. However, the band structure of the surface states is not strongly modified since they are below or above the Fermi level. In \cite{SM} we also present results for the other (subleading) chiral SC states.

\begin{figure}
\includegraphics[width=\columnwidth]{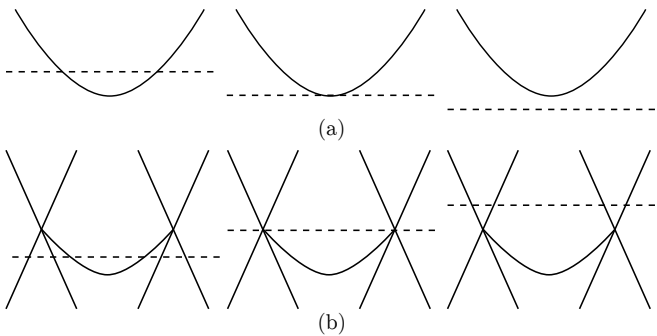}
\caption{Panel (a): A process to deform the weak pairing phase into strong pairing phase in 2D $p+ip$ SC, where the dashed lines are the chemical potential. It necessarily involves a gap closing at ${\bf k}=0$, where the SC order parameter is forced to vanish by fermionic statistics. Panel (b): The corresponding process for the drumhead surface SC, shown at a given $\theta$. No gap closing is involved with SC order turned on. In the last figure the drumhead dispersion can be further distorted.}
\label{gapclosing}
\end{figure}

{\it Topology of chiral surface SC state.---}We now address the topological properties of the chiral state ${{\bf d}}=d e^{i\theta}\hat y$, for which both the bulk and the surface are fully gapped.
 The topological properties of the bulk SC state has been analyzed in Ref. \cite{WeylLoopSC}, which used a homotopy argument to identify this state as a ``meron superconductor". Here we concentrate on the surface, which is a $p+ip$ pairing state. Such a state has been proposed for the quantum Hall state at $\nu=5/2$~\cite{moore-read, Anyon2}  and for the unconventional superconductor Sr$_2$RuO$_4$~\cite{SRO1,SRO2,SRO3}.
In the BdG Hamiltonian of a 2D $p+ip$ SC, one can define a homotopy group $\pi_2(S^2)$ from the 2D Brillouin zone to the Nambu spinor space, whose $z$-component is the normal part of the Hamiltonian. At TR invariant points of the BZ, the superconducting order parameter vanishes by fermionic statistics, and the Nambu spinor at these points is forced to point toward the north or south pole. Thus the winding (Chern) number is 1 if the BZ covers both poles, and  is 0 if it only cover one pole. The nontrivial case is with Chern number 1  corresponds to the case where the normal state has a Fermi surface, also known as the weak phase because the SC in this case is a weak-coupling instability in 2D. As a consequence of the topology, in the weak phase, it is well known that there exist MZMs bound at the core of a vortex of the pairing order parameter. The trivial strong phase, without a Fermi surface, requires a strong interaction to become paired on its own, and hence the name. To deform the strong phase Hamiltonian to the weak phase, the normal state energy $\epsilon({\bf k}=0)$ changes sign, and the gap closes in this process [see Fig.\ \ref{gapclosing}(a)], signifying a change in topology.

\begin{figure}
\includegraphics[width=\columnwidth]{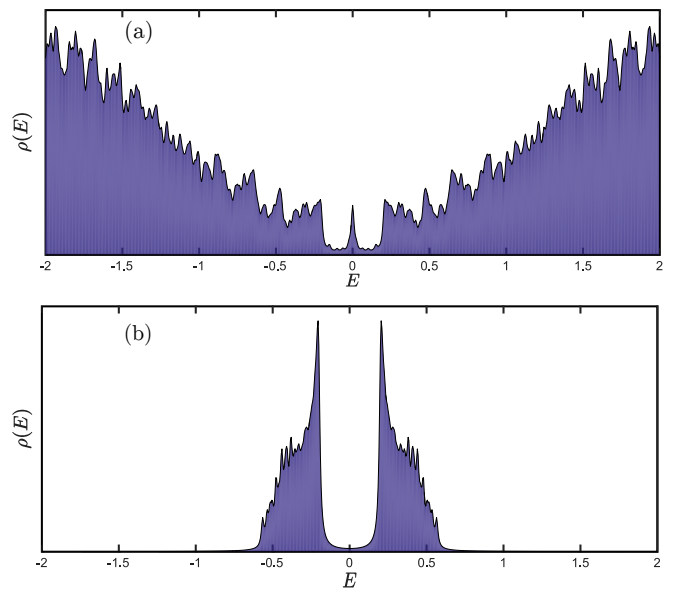}
\caption{The density of state plots for a Weyl loop chiral superconductor [Panel (a)] and a 2D $p+ip$ SC with a flat band dispersion [Panel (b)], both in the ``strong" pairing phase and with a vortex configuration. The spikes at high energies are due to finite-size effects (the system size is 
$20\times20\times 10$ for four bands).
For both Hamiltonians, there exists a bulk gap at $\mu=0.2$ (in unit of hopping parameters), however the clear distinction is that there exist in-gap states for the Weyl loop surface SC. }
\label{dos}
\end{figure}

For the surface SC states in our case, we {\it cannot} define the homotopy $\pi_2(S^2)$, because the surface bands do not traverse the full Brillouin zone. Despite the lack of a well-defined topological invariant, we can establish an analogy with the 2D $p+ip$ state. At first glance, it is tempting to classify the case with a surface FS as topological and the case without a surface FS as trivial. However, a careful examination of the band structure reveals that the  ``weak case"  {\it can} in fact be continuously deformed into the ``strong case" (note that despite the name, the SC order is from a weak-coupling instability in the bulk)  without closing the gap. We illustrate such a process in Fig. \ref{dos}(b). This implies these two cases are topologically equivalent and for both there should  exist MZMs when vortex lines enter and exit the bulk~\cite{footnote2}. However, the above analysis 
does not include the vortex line, which traps bound states. These vortex line states form a 1D band with a ``minigap".
If the minigap closes during this process then this could eliminate the MZMs (see e.g.~\cite{hosur-vishwanath}). We show in \cite{SM} that for our case minigap remains {\it open}.

To see how to obtain the MZM, we first look at a 2D $p+ip$ SC.
We approximate the lattice model with a Dirac fermion at the $\Gamma$ point,
${h}=\Delta(k_x\tau^x+ k_y\tau^y)-\mu\tau^z$,
where the $\tau$ matrices are in Nambu space and $\Delta$ is the SC order parameter, which has a vortex configuration. However, at this level, there is no distinction between the weak and strong phases. Moreover, the wave function for the bound state diverges as $r\to0$. A closer look shows that in the vortex core $r\sim0$ the system is in the normal state, and the wave function at such a short distance depends on the large momentum behavior of $\epsilon({\bf k})$, which is different for the weak and strong phases. One can then show that only for the dispersion of the weak phase can the wave function at small and large $r$ can be matched smoothly~\cite{kvorning,SM}.

However, the situation is different for our case. The surface drumhead band is not a standalone 2D system, and it only exists in a small momentum range around a TR invariant point. In the absence of large momentum states in the surface band, all the surface state wave functions are ``smoothed out" at short distances, eliminating the singularity at $r=0$~\cite{SM}. As a result the surface SC should be topological with vortex core MZMs independent of the drumhead band dispersion~\cite{footnote3}.  To check this logic, we numerically solved and compared the energy spectrum for a Weyl loop SC and 2D $p+ip$ SC, both in the ``strong" phase, and we found that only in the former case does there exist an in-gap modes with wave function peaked at the vortex. In Fig.\ \ref{dos} we show the comparative density of state plots indicating that surface SC in the doped Weyl loop semimetal is topological (in the sense that vortices trap MZM when they enter/exit the bulk) even when the drumhead lies entirely above/below the chemical potential. {We also verified that the same results hold for the weak phase of the Weyl loop SC.}

In conclusion, we have shown that in a doped Weyl loop material host entirely new types of topological surface superconductivity. For the fully gapped chiral SC state the surface surface traps MZM when vortex lines enter/exit the bulk, {\it independent} of the drumhead dispersion. Even for a ``conventional" $s$-wave state, there exists a flat band of neutral Majorana bound states on each surface at the projected location of the FS.

\acknowledgements{We acknowledge useful discussions with T. Kvorning, L. Santos, M. Stone and C. Wang. We also thank Yang-Zhi Chou, Shouvik Sur and especially Titus Neupert for feedback on the manuscript. This work was supported by the Gordon and Betty Moore Foundation's EPiQS Initiative through Grant No. GBMF4305 at the University of Illinois (Y.W.). Y.W. acknowledges support by the 2016 Boulder Summer School for Condensed Matter and Materials Physics through NSF grant DMR-13001648, where the initial ideas of this project were conceived.}

%merlin.mbs apsrev4-1.bst 2010-07-25 4.21a (PWD, AO, DPC) hacked
%Control: key (0)
%Control: author (8) initials jnrlst
%Control: editor formatted (1) identically to author
%Control: production of article title (-1) disabled
%Control: page (0) single
%Control: year (1) truncated
%Control: production of eprint (0) enabled
%

%%%%%%%% commands for reset equation, figure and refs numbering in supplementary material
\renewcommand{\theequation}{S\arabic{equation}}
\renewcommand{\thefigure}{S\arabic{figure}}
\renewcommand{\bibnumfmt}[1]{[S#1]}
\renewcommand{\citenumfont}[1]{S#1}
%%%%%%%%%%%%%%%%%%%%%%%%%%%%%%

\onecolumngrid

\newpage
\centerline{\large{\bf Supplemental Material}}

\section{I.~~~Surface states for other chiral SC orders}
For ${\tilde {\bf d}}=d\hat z$, the superconducting bulk has line nodes. However, on the surface $\chi_z^s(\varphi)e^{i\theta}=(\xi^s)^\dagger\sigma^x(\xi^s)^*e^{i\theta}=0$, where $\xi^s\equiv (1,0)^T$.
Thus the surface remains gapless and metallic. On the other hand, for ${\tilde {\bf d}}=d\hat x$,  the bulk is not gapped by SC order, but on the surface we have
$\chi_x^s(\varphi)e^{i\theta}=-(\xi^s)^\dagger\sigma^z(\xi^s)^*e^{i\theta}=-e^{i\theta}$,
which is a fully gapped state with a $p+ip$ symmetry. Therefore, in this case the bulk states remain metallic and we have a pure surface superconducting state.

\section{II.~~~Majorana flat band in the $s$-wave state}
In this section we show that each line node in the $s$-wave state carries a Berry flux $\pi$ around it. As we argued in the main text, this signifies a change in topology for the one-dimensional (1D) subsystems in $z$ across the line node. To see this, we linearize the BdG Hamiltonian in the vicinity of the line node (say for $\theta=0$ and $\varphi=0$)
\begin{align}
\tilde h ({\bf  k})=k_z\tau_y+(k_x-k_F)\tau_z,
\end{align}
where we have used the fact that $\varphi$ direction is along $z$ at this point, and $k_F$ is the radius of this nodal line. This dispersion describes a massless  two-dimensional (2D) Dirac fermion, and around the Dirac point (i.e. the line node), there is a Berry phase of $\pi$, since the Nambu pseudospin winds by $2\pi$.

The detailed calculation to show that in the $k$-space region bound by the two line nodes the 1D superconductor in $z$ carries Majorana end states is standard~\cite{kitaev-chain-s}, and we do not show it here.

\section{III.~~~Majorana zero modes in the chiral SC state}
In this Section, we first review the analysis of Majorana zero modes for a (2D) $p+ip$ superconductor. We show in detail how the boundary condition for the bound state wave function places a constraint on the normal state fermionic dispersion. We then discuss how the $p+ip$ superconductivity (SC) on the surface drumhead band of the doped Weyl loop semimetal can elude this requirement. We also elaborate on some details of the adiabatic continuation process used in the main text, in particular for the vortex line band and its ``minigap". As the analytic argument and numerical results show, the existence of Majorana mode for the $p+ip$ on the drumhead band indeed does not depend on its normal state dispersion.

\subsection{A.~~~Majorana zero modes in a 2D $p+ip$ SC}
To better understand the existence of Majorana zero modes of the surface drumhead SC, it is helpful to first analyze the case of a true 2D $p+ip$ SC.
Using a Chern number argument, it is well-known~\cite{Anyon2-s,AnyonReview1-s} that only for the cases with a normal state Fermi surface (FS), i.e., the weak pairing phase, the $p+ip$ state is topologically nontrivial. To connect adiabatically from the trivial case (the strong pairing phase, without a normal state FS) to the nontrivial case, the energy gap of the bulk Bogoliubov-deGennes (BdG) Hamiltonian necessarily closes during this process. However, what is less well-known is how the distinction between the two phases emerges from a direct calculation of the Majorana zero modes bound at the vortex cores. In many references~\cite{bernevig-book-s, AnyonReview1-s}, once the pairing state is established to be topological, the Majorana zero modes are obtained using a Dirac-form BdG Hamiltonian [Eq.\ (8) of the main text] by neglecting the fermionic dispersion. However, the Dirac-form Hamiltonian alone does not distinguish between the strong and weak pairing phases. In other words, only at the level of Dirac Hamiltonian, one seems to obtain a vortex-core zero mode for both topological and trivial phases. To see the topological distinction between the strong and weak pairing phases one needs to go beyond the Dirac-form Hamiltonian~\cite{kvorning-s, luiz-dispersion-s}.

We now present the detailed analysis of  the Majorana zero mode in a spinless $p+ip$ superconductor with a normal state dispersion $\hat\epsilon=-{\nabla^2}/{2m}-\mu$ (where we take $\mu>0$) in the presence of a vortex at $r=0$. 

One particular complication to the calculation is that the SC order parameter depends both on spatial coordinates (because of the vortex) and on momentum (because of the $p$-wave symmetry), which do not commute. The standard way in literature~\cite{Anyon3,kvorning-s,luiz-dispersion-s} is to take the anti-commutator of the coordinate-dependent part and the momentum-dependent part in the first-quantized BdG Hamiltonian. However it is more transparent to formulate the problem in the second-quantized language, and we shall see that this complication is automatically handled by fermionic statistics. The second-quantized Hamiltonian $H=\int d^2r \mathcal{H}(\bf r)$ is given by
\begin{align}
\mathcal{H}(\r)=&\frac{1}{2}\[\psi^\dagger(\r) \hat\epsilon \psi(\r) - \psi(\r) \hat\epsilon \psi^\dagger(\r) + \Delta (r) e^{i\theta} \psi^\dagger (\r)(\del_x+i\del_y) \psi^\dagger(\r) - \Delta (r) e^{-i\theta} \psi(\r) (\del_x -i\del_y) \psi(\r)   \]\nonumber\\
=&\frac{1}{2}\[\psi^\dagger(\r) \hat\epsilon \psi(\r) -\psi(\r) \hat\epsilon \psi^\dagger(\r) + \Delta (r) e^{2i\theta} \psi^\dagger (\r)\(\frac{\del}{\del r}+\frac{i}{r}\frac\del{\del\theta}\) \psi^\dagger(\r) - \Delta (r) e^{-2i\theta} \psi(\r) \(\frac{\del}{\del r}-\frac{i}{r}\frac\del{\del\theta}\) \psi(\r)   \],
\end{align}
where $\psi$ is a spinless fermion field, $r,\theta$ are polar coordinates, and in the second line we have used the relation $\del_x+i\del_y\equiv e^{i\theta} (\del_r + i\del_\theta/r)$. Note that the minus sign in the last term does {\it not} come from integration by parts, but from fermionic statistics. In the second-quantized language, the Majorana zero mode condition indicates the operator relation
\begin{align}
[H,\gamma] = 0,
\label{zero}
\end{align}
where we use the ansatz 
\begin{align}
\gamma=\int d^2r\  g(r)\[ f_1(\theta) \psi^\dagger (\r) + f_2 (\theta) \psi(\r)\].
\end{align}
Assuming $g$ and $f$'s are slowly varying functions, we can keep in the Hamiltonian only the lowest order gradient terms and take $\hat \epsilon=-\mu$.  Applying Eq.\ \eqref{zero} using the identity $[ab,c]=a\{b,c\}-\{a,c\}b$ then yields
\begin{align}
-\mu g(r) f_1(\theta) + \Delta(r) e^{2i\theta} \( \frac\del{\del r} - \frac{1}{r} +\frac{1}{2}\frac{\del \log \Delta(r)}{\del r} + \frac{i}{r}\frac{\del}{\del\theta}\) g(r) f_2(\theta)=&0, \nonumber\\
\mu g(r) f_2(\theta) - \Delta(r) e^{-2i\theta} \( \frac\del{\del r} - \frac{1}{r} +\frac{1}{2}\frac{\del \log \Delta(r)}{\del r} - \frac{i}{r}\frac{\del}{\del\theta}\) g(r) f_1(\theta)=&0.
\label{bdg}
\end{align}
In obtaining \eqref{bdg} we have integrated by parts in the second term of each line. These ``Schr\"odinger" equations for the Majorana wave function is indeed the same with what one would get using a first-quantized formalism while symmetrizing the $\r$ and ${\bf k}$ dependence.

The solution of this set of equations is given by
\begin{align}
f_1(&\theta)=e^{i\theta},~~~f_2(\theta)=-e^{-i\theta},\nonumber\\
g(r)=&\frac1{\sqrt{\Delta{(r)}}}\exp\[-\int^r \frac{\mu}{\Delta(r')} dr'\].
\label{wf}
\end{align}
For positive $\mu$ and $\Delta$ we can verify that the wave function given by this form is concentrated near $r=0$. For other signs of $\mu$ and $\Delta$, as well as for an antivortex with $\Delta(\r)\equiv \Delta(r) e^{-i\theta}$, a zero-mode wave function concentrated near $r=0$ can be similarly obtained.

Let us model a vortex configuration by $\Delta(\r)\equiv \Delta(r) e^{i\theta}$, where 
\begin{align}
\Delta(r)=\begin{cases} \Delta_0~r/R, ~~~r<R,\nonumber\\
\Delta_0.\end{cases}
\end{align}
Then the radial component of the Majorana mode wave function given by \eqref{wf} is 
\begin{align}
g(r)\sim\begin{cases} r^{-\mu R/2\Delta_0 -1/2}, ~~~&r<R,\nonumber\\
\exp\[-\frac{\mu r}{2\Delta_0}\],~~~& r>R,\end{cases}
\end{align}
which is a power law decay at short distances and an exponential decay at larger distances\footnote{Note that no extra care is needed to connect these two wave functions at $r=R$. They both come from Eq.\ \eqref{wf}, which applies well even when $\Delta(r)$ is not smooth at $r=R$.}. 

At first look this indeed describes a bound state at $r=0$. However, the wave function is singular at $r=0$ and is not normalizable for sufficiently small $\Delta_0$\footnote{Recall that on the FS the SC gap is $\Delta_0k_F\sim \Delta_0/a_0$, where $a_0$ is the lattice constant; even if we take $\Delta_0k_F\sim \mu$ the wave function has a power law divergence with a very large exponent $\sim R/a_0$.}. To resolve this issue, it is necessary to treat the region near $r=0$ more carefully~\cite{kvorning-s}. First, in this region the wave function is no longer smooth, and thus one need to restore the gradient terms in $\hat\epsilon$, i.e.\ fermionic dispersion. Second, in this region the SC order parameter vanishes and the system is essentially in the normal state. Therefore, in the vicinity of $r=0$, the problem reduces to solving for a  zero-energy bound state for a free fermion dispersion $\hat \epsilon = -\nabla^2/2m-\mu$ with angular momentum $\ell=\pm 1$ [see Eq.\ \eqref{wf}]. 
The Schr\"odinger equation for the radial wave function $\tilde g(r)$ with both $\ell=\pm1$ is a standard Bessel equation,
\begin{align}
\frac{1}{r}\frac{d}{dr}\[r\frac{\tilde g(r)}{dr}\]+ \(2m\mu-\frac{1}{r^2}\)\tilde g(r)=0. 
\label{sch}
\end{align} 
The solution only exists for $\mu>0$ (the weak pairing phase), and is given by the Bessel function 
\begin{align}
\tilde g(r)=J_{1}(\sqrt{2m\mu}r),
\end{align}
which is well-defined at $r=0$. \footnote{Had we used an anti-vortex configuration for $\Delta(\r)$, the solution in this region would have been $J_0(\sqrt{2m\mu}r)$.} This form of $\tilde g(r)$ is to be connected with the power-law form of $g(r)$ in Eq.\ \eqref{wf} at $r\sim 1/\sqrt{2m\mu}$, beyond which fermionic dispersion can be neglected. For the strong pairing phase $\mu<0$, such a solution to \eqref{sch} does not exist at $r\sim 0$ and hence the Majorana operator $\gamma$ is ill-defined.

Even without any of the technical details above, since $-\nabla^2/2m$ is a positive-definite operator, it is clear that $\hat\epsilon$ has an zero-energy eigenstate only for $\mu>0$.  This is precisely the condition for a normal state Fermi surface, i.e., the weak pairing phase. The solution $\tilde g(r)$ at $r\sim0$ above is nothing but a linear combination of all plane waves for FS fermions. We can thus unambiguously establish the weak pairing phase 
as the ``topological" phase. 

\begin{figure}
\includegraphics[width=0.3\columnwidth]{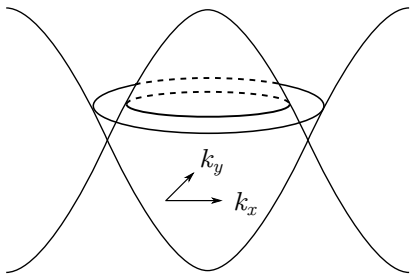}
\caption{The fermionic dispersion and two 2D Fermi surfaces at $k_z=0$.}
\label{circles}
\end{figure}

\subsection{B.~~~Majorana zero modes in the drumhead SC}
\subsubsection{1.~~~How the drumhead SC eludes the boundary condition at $r=0$ }
What we have learned in the previous Subsection is that, the identification of the weak pairing phase as the only topological phase relies crucially on the small-distance and large-momentum behavior of the wave function.  However, as we stated in the main text, for the surface drumhead band, there is no large momentum component -- the available momentum range is confined by the projection of the line node. As such, for wave functions localized on the surface, there is no singularity at $r=0$, which would have dictated a FS for the drumhead band (of course, there is a FS at large momentum in the {\it bulk}). Thus, we expect that the drumhead SC to be topological, no matter what the drumhead dispersion is. This is consistent with the fact that tuning between ``weak pairing phase" and the ``strong pairing phase" of the drumhead SC does not involve closing of the gap (more details below), as well as the numerical results we presented in the main text.

\subsubsection{2.~~~Fate of the minigap of the vortex line band during the adiabatic continuation}
We analyze the vortex line band formed by vortex core bound states. To obtain its dispersion in $k_z$, let us restore translational symmetry in $z$. Our goal is to show the minigap of this band remains open throughout the process described by Fig.\ 2 of the main text. 

The bulk states at every $k_z$ can be treated as a 2D SC in the class $D$, and the topology is characterized by a Chern number. For example, at $k_z=0$ the 2D subsystem has two FS's (two concentric circles shown in  Fig.\ \ref{circles}), which are the cross-section of the torus FS. In the fully-gapped chiral SC state, both Fermi surfaces are spin polarized by a $p+ip$ SC order. However, since one of the FS is electron-like and the other one is hole-like, the $p+ip$ order parameters on the two FS's contribute Chern numbers of 1 and $-1$ respectively. The total Chern number of the 2D SC is then zero and it is trivial. Therefore, with the insertion of a vortex line, which winds the SC order parameter in $xy$-real space, there is {\it no} zero-energy bound states at the vortex core. This means that the vortex line band, formed by the vortex bound states, is {\it gapped} at $k_z=0$.

One can similarly show the vortex line band is gapped at all $k_z$. This is because the Chern number for each $k_z$ has to be the same unless there is a Weyl point at some $k_z$, which our SC band structure clearly does not have. Therefore the $k$-space slices of the bulk SC at all $k_z$ are topologically trivial, and hence the vortex bound states at all $k_z$'s are gapped. This is consistent with the statement in Ref.\ \onlinecite{WeylLoopSC-s} that, despite the highly nontrivial band structure and SC order, there exists no well-defined topological invariant in the bulk.

Note that this conclusion applies to {\it all} the cases during the process in Fig.\ 2 of the main text. For example, at $k_z=0$, all this process does is switch the two Fermi surfaces in size. However, the bulk SC at $k_z=0$ remains topologically trivial. The same holds for other values of $k_z$. Therefore, the vortex line band remains gapped during the process depicted in the main text. 

From Fig.\ \ref{circles} we also see that the true critical case is when, upon a large doping, one of the two FS's shrinks to zero size. Beyond this point, at $k_z=0$ there is only one FS with a $p+ip$ pairing, which has a Chern number of 1, and the vortex line band is gapless. For other values of $k_z$ the Chern number is 1 until it hits a Weyl point. In this case the 3D FS is ``inflated" into a sphere, and the SC state is nothing but the spinless analog of the $A$-phase of $^3$He. The Weyl point is just the north or south pole of the FS, which remains gapless in the superconducting state. However, this is well beyond the situation we are interested in, i.e., when the bulk FS is a torus.

Finally, from this point of view, the presence of Majorana modes at the ends of the gapped vortex line indicates that it is in fact a 1D topological SC~\cite{kitaev-chain-s}. However, a direct evaluation of the $\mathbb{Z}_2$ topological invariant for this band is difficult as it requires the detailed 1D band structure, which we will not pursue here.


\begin{thebibliography}{44}
\bibitem {footnote1}{See however Ref.\
  \cite{Roy} for a discussion of proximity effects in Dirac loop materials,
  which differ from the Weyl loop materials discussed here by possessing an
  additional spin degeneracy.}
\bibitem {footnote2}%
  {Note that this
  conclusion would not hold for systems with two nodal lines that span the
  zone, since in that case the surface `drumhead' would extend to the zone
  boundary, and the gap must vanish by fermion statistics when the chemical
  potential is at the zone boundary.}
\bibitem{footnote3} {Note that, due to
  the large density of states on the drumhead band, it is entirely possible
  that the surface band becomes superconducting at a higher temperature than
  the bulk does. However, in this case the MZM on the top and bottom surface
  can couple through the gapless bulk.}
\end{thebibliography}
\end{document}